\documentclass[cameraready]{Interspeech}

\title{Beyond WER: A Paired Acoustic Stress Test for Ambient Clinical Scribes}

\author[affiliation={1}]{Xiao-Hang}{Jiang}
\author[affiliation={1}]{Han-Jie}{Guo}
\author[affiliation={1}]{Ying-Si}{Liang}
\author[affiliation={1}, correspondingauthor]{Yang}{Ai}
\author[affiliation={1}]{Zhen-Hua}{Ling}
\author[affiliation={2}]{Lei}{Jiang}
\author[affiliation={2}]{Zhi-Yang}{He}

\address{
    $^1$ University of Science and Technology of China, China \\
    $^2$ iFLYTEK Co., Ltd., China
}

\email{
\{jiang\_xiaohang, ghj2001, lys24204125\}@mail.ustc.edu.cn, \\
\{yangai, zhling\}@ustc.edu.cn, \{leijiang19, zyhe\}@iflytek.com
}

\usepackage{xcolor}

\usepackage{tabularx}
\usepackage{array}
\usepackage{makecell}

\usepackage{comment}

\begin{document}

\maketitle

\addtolength{\textfloatsep}{-0.5cm}
\addtolength{\dbltextfloatsep}{-0.5cm}

\begin{abstract}
Ambient clinical scribes increasingly combine Automatic Speech Recognition with Large Language Models to automate documentation. However, traditional metrics like Word Error Rate mask systemic safety degradation. We present a paired acoustic stress test to isolate the causal impact of noise on clinical reasoning. For the same dialogues, we inject diverse noise types while keeping the downstream model configuration frozen. Crucially, we uncover a dangerous disconnect between signal fidelity and clinical safety. Stationary ambient noise increased the Word Error Rate by a negligible 0.71 percentage points yet nearly doubled the rate of unsafe outputs. Our analysis reveals that minor acoustic perturbations can invert clinical meaning without substantially inflating error rates. Furthermore, we demonstrate a lightweight mitigation strategy that mitigates safety degradation under noisy conditions without requiring model fine tuning.
\end{abstract}

\section{Introduction}
Ambient clinical scribes, which cascade Automatic Speech Recognition (ASR) with Large Language Models (LLMs), are rapidly transforming healthcare documentation~\cite{coiera2019last, reiner2007radiology,rana2005voice,boland2007voice}. By unobtrusively recording clinician--patient dialogues and automating the generation of structured notes and decision support, these pipelines promise to alleviate severe documentation burdens and physician burnout~\cite{arndt2017tethered, tai2019physicians}. However, this shift fundamentally reallocates the burden of safety from human verification to algorithmic reliability. Cascading architectures are inherently vulnerable to error propagation: a seemingly minor misrecognition at the acoustic front-end can cascade into the LLM, precipitating severe diagnostic or triage errors~\cite{simonnet2017asr, chen2023improving}. Unlike human scribes, model errors can be fast, scalable, and deceptively well-formatted, introducing a high risk of silent failures that bypass standard clinical checks~\cite{finlayson2019adversarial}.

Traditionally, the robustness of the ASR front-end is evaluated using Word Error Rate (WER). Yet, recent literature highlights a critical misalignment: WER is fundamentally unaware of clinical semantics~\cite{ellis2026wer, lugosch2019speech}. The relationship between acoustic fidelity and downstream clinical integrity is highly non-linear. A single token perturbation—such as flipping a negation, altering a medication dosage, or swapping units (e.g., \emph{hours} vs.\ \emph{days})—can have a catastrophic clinical impact, whereas dropping conversational fillers is benign~\cite{kiefer2025instruction}. By treating all token errors as equally severe, aggregate transcript-level metrics create an \emph{illusion of accuracy}. They may suggest a system is robust while masking safety-critical semantic drift~\cite{binici2025medsage, wang2003word}. Consequently, the field lacks a systematic framework to answer a critical question: \emph{how} and \emph{why} do specific acoustic distortions drive downstream safety degradation?

To bridge this gap, we propose a \textbf{paired acoustic stress test} designed to evaluate the robustness degradation of ASR$\rightarrow$LLM clinical pipelines under realistic noise. Our core setup isolates the acoustic variable: if the exact same medical dialogue is heard under different noise conditions, does the clinical conclusion survive? For a given dialogue, we inject diverse background noises (e.g., multi-speaker babble, ambient clinic noise) at varying Signal-to-Noise Ratios (SNRs) to alter the ASR input~\cite{snyder2015musan, thiemann2013diverse}. Crucially, we keep the downstream LLM's configuration strictly frozen. This within-encounter design controls for case mix and generation stochasticity, allowing us to cleanly isolate the paired attribution of acoustic perturbations on downstream clinical reasoning~\cite{ribeiro2020beyond}.

Building on this controlled evaluation, this work shifts the paradigm from transcript exactness to clinical invariance. We introduce a novel taxonomy that explicitly links cause-side error trigger features (e.g., negation flips, number/unit inconsistencies, and non-speech pollution) to result-side safety endpoints (e.g., unexpected triage drift or omitted red flags). Table~\ref{tab:error_examples} provides concrete paired examples that motivate our cause-side trigger features and result-side safety endpoints. By systematically quantifying these degradation mechanisms, we demonstrate that specific semantic distortions, rather than aggregate WER, are the primary predictors of downstream safety failures. Ultimately, this framework suggests how ambient noise either structurally collapses or semantically traps LLM reasoning, exposing critical vulnerabilities in current clinical speech-to-text pipelines.

\begin{figure}[t]
  \centering
  \includegraphics[width=0.9\linewidth]{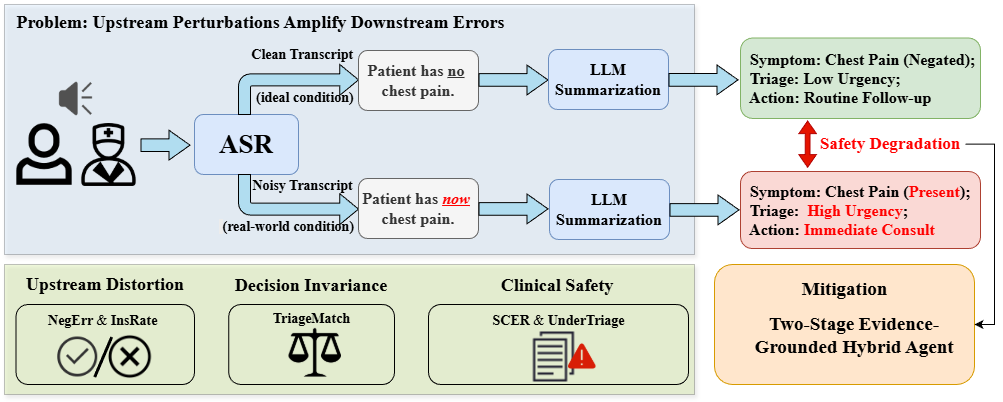}
  \caption{Overview of the paired acoustic stress test for clinical scribe pipelines.}
  \label{fig:overview}
\end{figure}

\begin{table*}[t]
\huge
\centering
\caption{Paired cause$\rightarrow$effect examples of ASR-induced safety degradation.}
\label{tab:error_examples}
\small
\resizebox{=0.93\textwidth}{!}{
\begin{tabular}{>{\raggedright\arraybackslash}p{3.4cm} >{\raggedright\arraybackslash}p{8.6cm} >{\raggedright\arraybackslash}p{6.8cm}}
\toprule
\textbf{ASR Error Type (Cause)} &
\textbf{Paired Transcript Example (Clean vs.\ Noisy)} &
\textbf{Downstream LLM Impact (Effect)} \\
\midrule

\textbf{Non-speech pollution} \newline
\textit{(background hallucination)} \newline
 &
\textbf{Clean:} ``\textbf{P:} I've had this dull ache in my right ankle...'' \newline
\textbf{Noisy:} ``...ache in my right ankle. \underline{\textit{Have you noticed blood in your stool at all? I take Metformin.}}'' &
\textbf{Spurious claim injection:}
Unsupported symptoms/medications are added to the structured note, which can inflate severity and introduce false red-flag mentions, increasing over-triage risk. \\
\midrule

\textbf{Critical number / unit conflict} \newline
\textit{(functional capacity distortion)} 
 &
\textbf{Clean:} ``\textbf{P:} I can walk about \underline{\textbf{3 or 4 flights of stairs}} before feeling breathless.'' \newline
\textbf{Noisy:} ``I can walk about \underline{\textbf{three or four steps}} before feeling breathless.'' &
\textbf{Severity amplification:}
A small number--unit swap converts mild limitation into near-complete exertional intolerance, shifting extracted functional status and potentially flipping triage and recommendations. \\
\midrule

\textbf{Negation loss / flip} \newline
\textit{(semantic polarity shift)} \newline
 &
\textbf{Clean:} ``\textbf{P:} It looks a little red and tender, but I \underline{\textbf{don't}} see that larger circular rash.'' \newline
\textbf{Noisy:} ``It looks a little red and tender, but I \underline{\textbf{[omitted]}} see that larger circular rash.'' &
\textbf{False positive finding:}
Dropping a single negation reverses a clinically meaningful statement, causing the structured note to record a symptom that was explicitly denied and altering downstream reasoning. \\
\midrule

\textbf{Temporal distortion} \newline
\textit{(acuity shift)} \newline
 &
\textbf{Clean:} ``\textbf{P:} The pain has been really intense for the last \underline{\textbf{two hours}}.'' \newline
\textbf{Noisy:} ``The pain has been really intense for the last \underline{\textbf{two days}}.'' &
\textbf{Acuity misinterpretation:}
Time-scale corruption changes perceived onset/acuity and can downgrade urgency, contributing to under-triage risk and missing time-sensitive evaluation. \\
\bottomrule
\end{tabular}
}
\end{table*}

\section{Proposed Method}
\label{sec:Methods}



\subsection{Overview}
\label{sec:method_overview}
As shown in Fig.~\ref{fig:overview}, we design a \textbf{controlled, paired acoustic stress test} to systematically attribute downstream clinical errors to upstream acoustic artifacts.
Formally, our evaluation pipeline processes an audio signal through an ASR front-end to generate a transcript $\mathbf{x}$, which is subsequently mapped into a structured clinical object
$y = \mathrm{LLM}_{\theta}(I, S, \mathbf{x}; \phi)$,
where $\theta$ denotes the fixed model parameters, $I$ is the system instruction prompt, $S$ is the fixed output schema/constraint specification, and $\phi$ denotes the decoding configuration.

Under our within-case design, each underlying clinical dialogue yields a clean baseline audio alongside multiple perturbed counterparts.
Crucially, we hold the downstream configuration fixed across all paired conditions, using identical $I$, $S$, and $\phi$ (with deterministic decoding).
This protocol disentangles acoustic sensitivity from generation stochasticity, ensuring that observed semantic drift in $y$ is attributable to the noise-induced perturbations in $\mathbf{x}$.

\subsection{The OSCE Evaluation Framework}
\label{sec:osce_framework}
To ensure clinical relevance without compromising patient privacy, our framework leverages the Objective Structured Clinical Examination (OSCE)  format\cite{harden1975assessment}.
OSCE is a widely used methodology in medical education where clinicians interact with standardized patients (actors) to simulate specific pathologies \cite{barrows1993overview}.
Beyond privacy compliance, this format is critical for safety benchmarking because scenarios are engineered to contain dense, specific safety cues (e.g., red flags, allergies) embedded within standard clinical workflows.
Unlike unstructured "in-the-wild" recordings where ground truth is often ambiguous, OSCE scenarios provide a deterministic reference, allowing us to rigorously quantify the omission of safety-critical claims under acoustic stress.

\subsection{Acoustic Interference Taxonomy}
\label{sec:noise_taxonomy}
To model realistic healthcare environments, we categorize acoustic artifacts into two distinct failure modes based on their stationarity and semantic content:
\begin{enumerate}
    \item \textbf{Stationary Ambient Interference:} Continuous, non-vocal background noise (e.g., HVAC hum, medical equipment drone) typical of consultation rooms. This primarily degrades ASR via spectral masking.
    \item \textbf{Non-Stationary Semantic Interference:} Overlapping human speech (e.g., multi-speaker babble in waiting rooms). This poses a severe safety risk as background voices actively compete with the primary dialogue \cite{brungart2001informational}, potentially inducing "hallucinated" medical entities (e.g., extracting a medication name spoken by a background speaker).
\end{enumerate}

\subsection{Condition Construction}
\label{sec:condition_construction}
To construct the paired evaluation set, all clean audio recordings are downmixed to mono, amplitude-normalized, and resampled to the same sample rate. 
For each clinical dialogue, we deterministically sample a continuous noise segment from a target acoustic corpus to match the dialogue's duration. 
We then mix the clean speech with the sampled noise at varying Signal-to-Noise Ratios (SNRs) by applying standard RMS-based energy scaling \cite{ko2017study}, which is computed solely over non-silent speech segments to prevent long-pause bias.
This procedure yields a robust dataset where every clean baseline transcript has paired perturbed counterparts, strictly controlling for patient history and clinical case mix.

\vspace{-1mm}
\section{Experiment}
\label{sec:experiments}

\subsection{Dataset Setup}
\label{sec:datasets}
\noindent\textbf{Clinical Corpus.}
We instantiate our OSCE framework using the open-source dataset by Fareez et al.~\cite{fareez2022dataset}. 
This corpus comprises \textbf{272} English encounters (approximately 52 hours), covering a diverse case mix of five specialties: respiratory, cardiovascular, gastrointestinal, musculoskeletal, and dermatological. All recordings are devoid of Protected Health Information (PHI).

\noindent\textbf{Noise Sources.}
We employ two standard benchmarks to represent our acoustic taxonomy:
(1) \textbf{DEMAND}~\cite{thiemann2013diverse} (\textit{Office, Cafeteria, Traffic} subsets) represents Stationary Ambient Interference.
(2) \textbf{MUSAN}~\cite{snyder2015musan} (\textit{Speech} subset) represents Non-Stationary Semantic Interference.
We evaluate at SNRs of $\{15, 10, 5\}$~dB to simulate environments ranging from quiet clinics to chaotic emergency departments.

\subsection{Pipeline Configuration}
\label{sec:system_setup}
\noindent\textbf{ASR Front-End.}
We employ \textsc{Whisper-large-v3}~\cite{radford2023robust} for transcription.
To mitigate decoding stochasticity and isolate acoustic artifacts as the primary driver of variation, we enforce deterministic greedy decoding (temperature=$0$) with a fixed 30-second window.

\noindent\textbf{Clinical Structuring.}
We utilize \textsc{Qwen3-235B-A22B-Instruct-2507} \cite{yang2025qwen3} (via an official API service) to map ASR transcripts into structured JSON documentation.
To enforce output fidelity, the system instruction imposes evidence-grounded constraints:
safety-critical assertions (e.g., allergies) must cite verbatim source spans to mitigate hallucination, and decision outputs are restricted to a fixed taxonomy (e.g., a 3-point triage scale) to enable rigorous quantitative comparison across conditions.

\subsection{Evaluation Paradigm}
\label{sec:gold_standard}
Clinical safety evaluation relies on the correctness of \emph{actionable semantics} rather than surface-form text overlap (e.g., ROUGE). 
To rigorously benchmark safety, we establish a clinician-audited gold reference via a model-assisted, human-verified workflow \cite{tu2025towards}.

For all $N{=}272$ encounters, we first employed \textsc{GPT-5.2} \cite{singh2025openai} (via an official API service) to decompose the clean human transcripts into a set of atomic clinical facts (e.g., specific symptoms, medication status, timeframes). 
Crucially, these candidate claims underwent a physician audit, where clinical experts corrected hallucinations and supplemented omitted details to ensure high clinical validity.
This process yielded a ground-truth set of semantic units for each encounter.
All downstream metrics compare the model's structured output against this physician-verified reference, isolating genuine semantic safety drift from benign phrasing variations.

\begin{table*}[t]
\centering
\caption{\textbf{Main results of the paired acoustic stress test.}}
\label{tab:mainstress_merged}
\resizebox{0.93\textwidth}{!}{
\begin{tabular}{l c c c c c c c c c}
\toprule
\textbf{Condition} &
\textbf{WER$\downarrow$} &
\textbf{InsRate$\downarrow$} &
\textbf{NegErr$\downarrow$} &
\textbf{ErrProp$\downarrow$} &
\textbf{TriageMatch$\uparrow$} &
\textbf{SCER$\downarrow$} &
\textbf{UnderTriage$\downarrow$} &
\textbf{Mean Score$\uparrow$} &
\textbf{Unsafe$\downarrow$} \\
& \scriptsize{(\%)} & \scriptsize{(\%)} & \scriptsize{(\%)} & \scriptsize{(new-err/100 $\Delta$ASR errs)} & \scriptsize{(\%)} & \scriptsize{(\%)} & \scriptsize{(\%)} & \scriptsize{(1--5 Scale)} & \scriptsize{(\%)} \\
\midrule
Reference (clean-ASR) & 16.54 & 2.99 & 19.12 & -- & 100.00 & 0.00 & 0.00 & 4.62 & 13.60 \\
\midrule
\multicolumn{10}{l}{\textit{\textbf{Non-Stationary Semantic Interference (MUSAN Speech)}}} \\
\hspace{3mm} 15 dB & 21.16 & 4.74 & 34.93 & 68.49 & 91.91 & 75.37 & 2.21 & 3.92 & 66.91 \\
\hspace{3mm} 10 dB & 30.00 & 8.29 & 37.50 & 25.31 & 91.54 & 83.46 & 3.31 & 3.74 & 78.68 \\ 
\hspace{3mm} 5 dB  & 54.68 & 14.55 & 51.10 & 10.53 & 81.99 & 92.28 & 7.72 & 3.41 & 91.54 \\
\midrule
\multicolumn{10}{l}{\textit{\textbf{Stationary Ambient Interference (DEMAND Office, Cafeteria, Traffic)}}} \\
\hspace{3mm} 15 dB & 17.25 & 2.48 & 20.22 & 272.83 & 93.01 & 44.12 & 3.31 & 4.45 & 27.21 \\
\hspace{3mm} 10 dB & 18.02 & 2.30 & 26.10 & 172.96 & 92.65 & 46.32 & 2.94 & 4.43 & 25.74 \\ 
\hspace{3mm} 5 dB  & 20.63 & 2.31 & 39.71 & 77.93 & 91.54 & 66.18 & 3.31 & 4.23 & 40.44 \\
\midrule
\multicolumn{10}{l}{\textbf{\textit{Mitigation Study (Agent-Based Framework)}}} \\
\hspace{3mm} Semantic 5 dB & 54.68 & 14.55 & 51.10 & 10.62 & 86.03 & 83.82 & 2.57 & 3.71 & 70.96 \\ 
\hspace{3mm} Ambient 5 dB & 20.63 & 2.31 & 39.71 & 78.82 & 92.28 & 51.84 & 1.10 & 4.36 & 33.46 \\ 
\bottomrule
\end{tabular}
}
\end{table*}
\subsection{Metrics}
\label{sec:metrics}

\subsubsection{Upstream Signal Fidelity (Cause-Side)}

\noindent\textbf{Word Error Rate (WER) \& Insertion Rate (InsRate).}
Let $\mathcal{D}$ be the evaluation dialogue set, indexed by $i$, with $|\mathcal{D}|=M$.
For each dialogue $i$, let $r_i$ be the human reference transcript, $a_i^{(0)}$ the ASR transcript from clean audio (clean-ASR baseline), and $a_i^{(c)}$ the ASR transcript under noisy condition $c$.
Let $S_i^{(c)}, D_i^{(c)}, I_i^{(c)}$ be the Levenshtein counts of substitutions, deletions, and insertions when aligning $a_i^{(c)}$ to $r_i$, and let $N_i$ be the number of words in $r_i$.
We report micro-averaged rates over tokens:
\begin{equation}
\mathrm{WER}(c) = 100 \cdot \frac{\sum_{i\in\mathcal{D}}\left(S_i^{(c)}+D_i^{(c)}+I_i^{(c)}\right)}{\sum_{i\in\mathcal{D}} N_i}.
\end{equation}
\begin{equation}
\mathrm{InsRate}(c) = 100 \cdot \frac{\sum_{i\in\mathcal{D}} I_i^{(c)}}{\sum_{i\in\mathcal{D}} N_i}.
\end{equation}
\noindent
InsRate is a lightweight proxy for additive transcript artifacts (e.g., repetitions, filler-like insertions, or noise-induced extra words), capturing general over-generation beyond substitutions/deletions.

\noindent\textbf{Negation Error Rate (NegErr).}
Let $\mathcal{N}(r_i, a_i^{(c)})\in\{0,1\}$ indicate whether any clinically relevant entity in dialogue $i$ exhibits a negation-status mismatch between $r_i$ and $a_i^{(c)}$ (e.g., \emph{no fever} vs. \emph{fever}).
\begin{equation}
\mathrm{NegErr}(c) = \frac{100}{M}\sum_{i\in\mathcal{D}} \mathcal{N}(r_i, a_i^{(c)}).
\end{equation}

\subsubsection{Downstream Robustness \& Safety (Result-Side)}

\noindent\textbf{Notation and evaluation scope.}
For each dialogue $i$ and condition $c$, the downstream pipeline produces a structured output that includes (i) an ordinal triage label and (ii) a set of atomic clinical claims / safety flags.
In our evaluation, these fields are present for every dialogue under every condition; therefore, all downstream metrics below are computed on the full set $\mathcal{D}$ (no instance filtering).

\noindent\textbf{Paired Triage Match (Invariance).}
Let $t_i^{(0)}$, $t_i^{(c)}$, and $t_i^{(\mathrm{gold})}$ be the triage labels predicted from the clean baseline, noisy condition $c$, and the gold reference, respectively.
We measure paired invariance as agreement with the clean baseline:
\begin{equation}
\mathrm{TriageMatch}(c)=\frac{1}{M}\sum_{i\in\mathcal{D}} \mathbf{1}\left[t_i^{(c)}=t_i^{(0)}\right].
\end{equation}

\noindent\textbf{Under-Triage Rate (Risk).}
Let $\mathrm{ord}(t)$ map a triage label to an ordinal severity rank (larger = more urgent).
Define $U_i^{(c)}=\mathbf{1}\left[\mathrm{ord}(t_i^{(c)}) < \mathrm{ord}(t_i^{(\mathrm{gold})})\right]$,
where $\mathbf{1}[\cdot]$ is the indicator function.
We report the rate of \emph{new} under-triage cases induced by noise:
\begin{equation}
\mathrm{UnderTriage}(c)=\frac{100}{M}\sum_{i\in\mathcal{D}}
\mathbf{1}\left[U_i^{(c)}=1 \wedge U_i^{(0)}=0\right].
\end{equation}

\noindent\textbf{Safety-Critical Error Rate (SCER).}
Let $G_i^{(c)}$ be the set of extracted safety flags under condition $c$, $G_i^{(\mathrm{gold})}$ the gold flags, and $\mathcal{G}_{\mathrm{crit}}$ the predefined safety-critical subset.
Define noise-induced critical \emph{false-positive} flags:
\[
\Delta G_i^{(c)} = \left(G_i^{(c)} \setminus G_i^{(\mathrm{gold})}\right)\cap \mathcal{G}_{\mathrm{crit}}.
\]
SCER measures whether noise introduces any \emph{new} critical false-positive flags relative to the clean baseline:
\begin{equation}
\mathrm{SCER}(c)=\frac{100}{M}\sum_{i\in\mathcal{D}}
\mathbf{1}\left[\left|\Delta G_i^{(c)} \setminus \Delta G_i^{(0)}\right|>0\right].
\end{equation}
\noindent
SCER targets noise-induced hallucinations (false positives), while safety risks arising from omissions (false negatives) are reflected by the complementary \textit{UnderTriage} metric and the holistic \textit{Unsafe Rate}.

\noindent\textbf{Error Propagation Rate (ErrProp).}
Let $C_i^{(c)}$ and $C_i^{(\mathrm{gold})}$ be the sets of atomic claims under condition $c$ and gold reference.
Let $\triangle$ denote symmetric difference.
Define claim error set $\mathrm{Err}_i^{(c)} = C_i^{(c)} \triangle C_i^{(\mathrm{gold})}$ and newly introduced claim errors
$\Delta \mathrm{Err}_i^{(c)} = \left|\mathrm{Err}_i^{(c)} \setminus \mathrm{Err}_i^{(0)}\right|$.
Let $E_i^{(c)} = S_i^{(c)}+D_i^{(c)}+I_i^{(c)}$ be the ASR word-error count under condition $c$ and
$\Delta E_i^{(c)}=\max\left(0, E_i^{(c)}-E_i^{(0)}\right)$ be additional ASR errors induced by noise.
Then:
\begin{equation}
\mathrm{ErrProp}(c)=100 \cdot \frac{\sum_{i\in\mathcal{D}} \Delta \mathrm{Err}_i^{(c)}}{\sum_{i\in\mathcal{D}} \Delta E_i^{(c)}}.
\end{equation}
\noindent
When the denominator is zero (i.e., no additional ASR errors relative to the baseline), we define $\mathrm{ErrProp}(c)=0$ for reporting.
ErrProp quantifies the marginal yield of \emph{new} gold-referenced claim errors per additional ASR word errors.

\subsubsection{Graded Clinical Consistency Scoring}
\label{sec:llm_judge}

We use \textsc{GPT-5.2} as a deterministic rubric-based judge, following G-Eval~\cite{liu2023g}, to assess clinical consistency between each generated note and the clinician-audited gold reference.
The judge is instructed to focus on clinically meaningful deviations (i.e., claim-level errors) rather than stylistic similarity; semantically equivalent paraphrases are not penalized.

We adopt a clinically grounded 1--5 rubric:
\begin{itemize}
    \item \textbf{5 (Clinically Equivalent):} No clinically meaningful deviations from gold on all safety-critical claims (e.g., red flags, triage-relevant cues, numbers/units/time, contraindications).
    \item \textbf{4 (Minor Non-critical Deviations):} Minor omissions or inaccuracies that do not affect clinical risk assessment or immediate decisions; safety-critical claims remain correct.
    \item \textbf{3 (Potentially Actionable Ambiguity):} Missing or unclear information that may require clarification, but does not constitute a clear high-risk error given available evidence.
    \item \textbf{2 (High-risk Deviation):} At least one major error relative to gold that can plausibly increase risk (e.g., omission of a documented red flag, incorrect number/unit/time, or incorrect triage-relevant interpretation).
    \item \textbf{1 (Critical Failure):} Safety-critical hallucination or a deviation that could directly lead to dangerous under-triage or harmful action.
\end{itemize}

Let $s_i^{(c)}\in\{1,\dots,5\}$ be the judge score for dialogue $i$ under condition $c$.
We report \textbf{Mean Score} $\frac{1}{M}\sum_{i\in\mathcal{D}} s_i^{(c)}$ and
\textbf{Unsafe Rate} $\frac{100}{M}\sum_{i\in\mathcal{D}}\mathbf{1}[s_i^{(c)}\le2]$.

\subsection{Mitigation: Evidence-Grounded Hybrid Agent}
To reduce noise-induced hallucinations at severe SNR (5\,dB), we introduce a lightweight \emph{evidence-grounded auditing} pipeline implemented with \textsc{Qwen3-235B-A22B-Instruct-2507} backbone.
The agent operates via two roles using a \textbf{dual-view context} (raw transcript for evidence, cleaned view for reasoning).
\textbf{Draft (Extractor)} generates the structured JSON and attaches \emph{verbatim} evidence quotes in \texttt{field\_evidence} for safety-relevant fields.
\textbf{Guard (Verifier)} audits the draft against the raw transcript; it is constrained to \emph{remove} unsupported content, preferring abstention when evidence is missing.

We further apply a deterministic \textbf{symbolic evidence filter}: (i) any quoted evidence that is not an exact substring of the raw transcript is discarded; and (ii) high-stakes fields are set to \texttt{null} if no surviving evidence remains.
We use temperature $=0$.
This design deliberately trades completeness for traceability, targeting reductions in safety-critical ``never-event'' errors (Unsafe/SCER) under heavy acoustic corruption.

\subsection{Results and Analysis}
\label{sec:results}

Table~\ref{tab:mainstress_merged} summarizes the paired acoustic stress test across 272 OSCE-style dialogues.
Holding the downstream pipeline fixed, we characterize failure modes across upstream ASR distortion, downstream invariance, and safety degradation.
Two key observations emerge: (i) transcript fidelity is a weak proxy for clinical safety, as the clean-ASR baseline yields a non-trivial Unsafe Rate; and (ii) noise type strongly shapes downstream errors.

\noindent\textbf{Upstream distortion signatures.}
Non-stationary semantic interference (MUSAN) produces an insertion-heavy transcript profile, with InsRate rising sharply as SNR decreases.
Conversely, stationary ambient noise (DEMAND) minimally impacts surface metrics (WER, InsRate) but substantially degrades NegErr.
This pattern is consistent with failures being driven by high-leverage semantic cues (e.g., negations), reinforcing that WER alone cannot anticipate safety outcomes.

\noindent\textbf{Safety degradation without invariance breaks.}
Under semantic interference, degradation is overt: invariance drops and safety failures become pervasive.
In contrast, ambient interference exhibits a silent failure mode: despite relatively stable triage outputs, safety violations increase markedly.
This gap highlights that self-consistency does not imply correctness; clinically meaningful errors can accumulate heavily even while the system remains stable with respect to its clean-ASR output.

\noindent\textbf{High-leverage error propagation.}
A notable spike occurs under mild ambient noise, indicating an unusually high marginal claim-error yield per additional ASR error.
As SNR decreases, ErrProp declines numerically as lower-impact word errors inflate the denominator. 
Although absolute safety risks generally trend upward with stronger noise, minor non-monotonic fluctuations under ambient interference are consistent with the effect depending on which critical utterances are acoustically perturbed, rather than being a purely linear function of SNR. 
Overall, small transcript perturbations can be disproportionately consequential at the claim level.

\noindent\textbf{Mitigation via evidence-grounding.}
Deploying the \emph{Evidence-Grounded Hybrid Agent} under severe 5\,dB conditions improves safety across both noise regimes.
By enforcing verbatim constraints, the agent reduces high-risk outcomes while leaving overall ErrProp largely unchanged.
This sustained ErrProp alongside reduced Unsafe and SCER rates indicates the symbolic filter disproportionately prunes unsupported, safety-critical ``never-event'' content via evidence-based abstention, while leaving benign claim variations relatively unaffected.

\section{Conclusion}
\label{sec:conclusion}

We presented a paired counterfactual noise stress test for ASR$\rightarrow$LLM clinical scribe pipelines, isolating how controlled acoustic perturbations propagate into downstream clinical-claim drift.
Across realistic noise families, we find that safety-relevant errors can increase even when transcript-level fidelity changes appear modest, highlighting that WER alone is an inadequate proxy for end-to-end clinical safety.
Different noise profiles induce distinct distortion patterns and downstream failure modes, motivating claim-aware evaluation.
Finally, our lightweight hybrid agent shows that symbolic evidence constraints can reduce high-risk hallucinations under severe acoustic corruption.

\section{Generative AI Use Disclosure}
During the preparation of this manuscript, the authors used ChatGPT 5.2 to polish the language and improve the flow of the text. After using this tool, the authors reviewed and edited the content as needed and take full responsibility for the final version of the manuscript.

\bibliographystyle{IEEEtran}
\bibliography{mybib}

@article{coiera2019last,
  title={The last mile: where artificial intelligence meets reality},
  author={Coiera, Enrico},
  journal={Journal of medical Internet research},
  volume={21},
  number={11},
  pages={e16323},
  year={2019},
  publisher={JMIR Publications Toronto, Canada}
}

@article{reiner2007radiology,
  title={Radiology reporting, past, present, and future: the radiologist’s perspective},
  author={Reiner, Bruce I and Knight, Nancy and Siegel, Eliot L},
  journal={Journal of the American College of Radiology},
  volume={4},
  number={5},
  pages={313--319},
  year={2007},
  publisher={Elsevier}
}

@article{rana2005voice,
  title={Voice recognition for radiology reporting: is it good enough?},
  author={Rana, DS and Hurst, G and Shepstone, L and Pilling, J and Cockburn, J and Crawford, M},
  journal={Clinical radiology},
  volume={60},
  number={11},
  pages={1205--1212},
  year={2005},
  publisher={Elsevier}
}

@article{boland2007voice,
  title={Voice recognition technology for radiology reporting: transforming the radiologist's value proposition},
  author={Boland, Giles WL},
  journal={Journal of the American College of Radiology},
  volume={4},
  number={12},
  pages={865--867},
  year={2007},
  publisher={Elsevier}
}

@article{arndt2017tethered,
  title={Tethered to the EHR: primary care physician workload assessment using EHR event log data and time-motion observations},
  author={Arndt, Brian G and Beasley, John W and Watkinson, Michelle D and Temte, Jonathan L and Tuan, Wen-Jan and Sinsky, Christine A and Gilchrist, Valerie J},
  journal={The Annals of Family Medicine},
  volume={15},
  number={5},
  pages={419--426},
  year={2017},
  publisher={The Annals of Family Medicine}
}

@inproceedings{simonnet2017asr,
  title={{ASR} error management for improving spoken language understanding},
  author={Simonnet, Edwin and Ghannay, Sahar and Camelin, Nathalie and Est{\`e}ve, Yannick and de Mori, Renato},
  booktitle={Interspeech 2017},
  year={2017}
}

@article{tai2019physicians,
  title={Physicians’ well-being linked to in-basket messages generated by algorithms in electronic health records},
  author={Tai-Seale, Ming and Dillon, Ellis C and Yang, Yan and Nordgren, Robert and Steinberg, Ruth L and Nauenberg, Teresa and Lee, Tim C and Meehan, Amy and Li, Jinnan and Chan, Albert Solomon and others},
  journal={Health affairs},
  volume={38},
  number={7},
  pages={1073--1078},
  year={2019}
}

@article{finlayson2019adversarial,
  title={Adversarial attacks on medical machine learning},
  author={Finlayson, Samuel G and Bowers, John D and Ito, Joichi and Zittrain, Jonathan L and Beam, Andrew L and Kohane, Isaac S},
  journal={Science},
  volume={363},
  number={6433},
  pages={1287--1289},
  year={2019},
  publisher={American Association for the Advancement of Science}
}

@inproceedings{ellis2026wer,
  title={{WER} is Unaware: Assessing How ASR Errors Distort Clinical Understanding in Patient Facing Dialogue},
  author={Ellis, Zachary and Joselowitz, Jared and Deo, Yash and He, Yajie and Kalygina, Anna and Higham, Aisling and Rahimzadeh, Mana and Jia, Yan and Habli, Ibrahim and Lim, Ernest},
  booktitle={Proceedings of the 13th International Workshop on Spoken Dialogue Systems Technology (IWSDS)},
  year={2026}
}

@inproceedings{ko2017study,
  title={A study on data augmentation of reverberant speech for robust speech recognition},
  author={Ko, Tom and Peddinti, Vijayaditya and Povey, Daniel and Seltzer, Michael L and Khudanpur, Sanjeev},
  booktitle={2017 IEEE international conference on acoustics, speech and signal processing (ICASSP)},
  pages={5220--5224},
  year={2017},
  organization={IEEE}
}

@article{tu2025towards,
  title={Towards conversational diagnostic artificial intelligence},
  author={Tu, Tao and Schaekermann, Mike and Palepu, Anil and Saab, Khaled and Freyberg, Jan and Tanno, Ryutaro and Wang, Amy and Li, Brenna and Amin, Mohamed and Cheng, Yong and others},
  journal={Nature},
  volume={642},
  number={8067},
  pages={442--450},
  year={2025},
  publisher={Nature Publishing Group UK London}
}

@article{singh2025openai,
  title={Openai gpt-5 system card},
  author={Singh, Aaditya and Fry, Adam and Perelman, Adam and Tart, Adam and Ganesh, Adi and El-Kishky, Ahmed and McLaughlin, Aidan and Low, Aiden and Ostrow, AJ and Ananthram, Akhila and others},
  journal={arXiv preprint arXiv:2601.03267},
  year={2025}
}

@article{barrows1993overview,
  title={An overview of the uses of standardized patients for teaching and evaluating clinical skills. AAMC},
  author={Barrows, Howard S},
  journal={Academic medicine},
  volume={68},
  number={6},
  pages={443--51},
  year={1993}
}

@article{yang2025qwen3,
  title={Qwen3 technical report},
  author={Yang, An and Li, Anfeng and Yang, Baosong and Zhang, Beichen and Hui, Binyuan and Zheng, Bo and Yu, Bowen and Gao, Chang and Huang, Chengen and Lv, Chenxu and others},
  journal={arXiv preprint arXiv:2505.09388},
  year={2025}
}

@article{harden1975assessment,
  title={Assessment of clinical competence using objective structured examination.},
  author={Harden, Ronald M and Stevenson, Mary and Downie, W Wilson and Wilson, GM},
  journal={Br Med J},
  volume={1},
  number={5955},
  pages={447--451},
  year={1975},
  publisher={British Medical Journal Publishing Group}
}

@inproceedings{ribeiro2020beyond,
  title={Beyond accuracy: Behavioral testing of NLP models with CheckList},
  author={Ribeiro, Marco Tulio and Wu, Tongshuang and Guestrin, Carlos and Singh, Sameer},
  booktitle={Proceedings of the 58th annual meeting of the association for computational linguistics},
  pages={4902--4912},
  year={2020}
}

@inproceedings{radford2023robust,
  title={Robust speech recognition via large-scale weak supervision},
  author={Radford, Alec and Kim, Jong Wook and Xu, Tao and Brockman, Greg and McLeavey, Christine and Sutskever, Ilya},
  booktitle={International conference on machine learning},
  pages={28492--28518},
  year={2023},
  organization={PMLR}
}

@article{fareez2022dataset,
  title={A dataset of simulated patient-physician medical interviews with a focus on respiratory cases},
  author={Fareez, Faiha and Parikh, Tishya and Wavell, Christopher and Shahab, Saba and Chevalier, Meghan and Good, Scott and De Blasi, Isabella and Rhouma, Rafik and McMahon, Christopher and Lam, Jean-Paul and others},
  journal={Scientific Data},
  volume={9},
  number={1},
  pages={313},
  year={2022},
  publisher={Nature Publishing Group UK London}
}

@article{brungart2001informational,
  title={Informational and energetic masking effects in the perception of two simultaneous talkers},
  author={Brungart, Douglas S},
  journal={The Journal of the Acoustical Society of America},
  volume={109},
  number={3},
  pages={1101--1109},
  year={2001},
  publisher={Acoustical Society of America}
}

@inproceedings{thiemann2013diverse,
  title={The diverse environments multi-channel acoustic noise database (demand): A database of multichannel environmental noise recordings},
  author={Thiemann, Joachim and Ito, Nobutaka and Vincent, Emmanuel},
  booktitle={Proceedings of Meetings on Acoustics},
  volume={19},
  number={1},
  pages={035081},
  year={2013},
  organization={Acoustical Society of America}
}

@article{snyder2015musan,
  title={Musan: A music, speech, and noise corpus},
  author={Snyder, David and Chen, Guoguo and Povey, Daniel},
  journal={arXiv preprint arXiv:1510.08484},
  year={2015}
}

@inproceedings{wang2003word,
  title={Is word error rate a good indicator for spoken language understanding accuracy},
  author={Wang, Ye-Yi and Acero, Alex and Chelba, Ciprian},
  booktitle={2003 IEEE workshop on automatic speech recognition and understanding (IEEE Cat. No. 03EX721)},
  pages={577--582},
  year={2003},
  organization={IEEE}
}

@inproceedings{binici2025medsage,
  title={{MEDSAGE}: enhancing robustness of medical dialogue summarization to ASR errors with llm-generated synthetic dialogues},
  author={Binici, Kuluhan and Kashyap, Abhinav Ramesh and Schlegel, Viktor and Liu, Andy T and Dwivedi, Vijay Prakash and Nguyen, Thanh-Tung and Gao, Xiaoxue and Chen, Nancy F and Winkler, Stefan},
  booktitle={Proceedings of the AAAI Conference on Artificial Intelligence},
  volume={39},
  number={22},
  pages={23496--23504},
  year={2025}
}

@inproceedings{kiefer2025instruction,
  title={Instruction-Tuning {LLaMA} for Synthetic Medical Note Generation in Swedish and English},
  author={Kiefer, Lotta and Alabi, Jesujoba and Vakili, Thomas and Dalianis, Hercules and Klakow, Dietrich},
  booktitle={Proceedings of the 15th International Conference on Recent Advances in Natural Language Processing-Natural Language Processing in the Generative AI Era},
  pages={557--566},
  year={2025}
}

@inproceedings{chen2023improving,
  title={Improving the robustness of summarization systems with dual augmentation},
  author={Chen, Xiuying and Long, Guodong and Tao, Chongyang and Li, Mingzhe and Gao, Xin and Zhang, Chengqi and Zhang, Xiangliang},
  booktitle={Proceedings of the 61st Annual Meeting of the Association for Computational Linguistics (Volume 1: Long Papers)},
  pages={6846--6857},
  year={2023}
}

@inproceedings{liu2023g,
  title={G-eval: NLG evaluation using gpt-4 with better human alignment},
  author={Liu, Yang and Iter, Dan and Xu, Yichong and Wang, Shuohang and Xu, Ruochen and Zhu, Chenguang},
  booktitle={Proceedings of the 2023 conference on empirical methods in natural language processing},
  pages={2511--2522},
  year={2023}
}

@inproceedings{lugosch2019speech,
  title={Speech Model Pre-{T}raining for End-to-{E}nd Spoken Language Understanding},
  author={Lugosch, Loren and Ravanelli, Mirco and Ignoto, Patrick and Tomar, Vikrant Singh and Bengio, Yoshua},
  booktitle={Interspeech},
  year={2019},
  organization={ISCA}
}

\end{document}